\documentclass[12pt]{iopart}
\usepackage{iopams}  
\usepackage{graphicx}
\newcommand{\dket}[1]{| #1 \rangle}
\newcommand{\dbra}[1]{\langle #1  |}
\begin{document}
\jl{9}

\title[Minimum error discrimination of Pauli channels]{Minimum error
discrimination of Pauli channels}

\author{M F Sacchi}

\address{QUIT, INFM and Dipartimento di Fisica ``A. Volta'',
  Universit\`a di Pavia, via A. Bassi 6, I-27100 Pavia, Italy}

\begin{abstract}
  We solve the problem of discriminating with minimum error
  probability two given Pauli channels. We show that, differently from
  the case of discrimination between unitary transformations, the use
  of entanglement with an ancillary system can strictly improve the
  discrimination, and any maximally entangled state allows to achieve
  the optimal discrimination. We also provide a simple necessary and
  sufficient condition in terms of the structure of the channels for
  which the ultimate minimum error probability can be achieved without
  entanglement assistance. When such a condition is satisfied, the
  optimal input state is simply an eigenstate of one of the Pauli
  matrices.
\end{abstract}


\section{Introduction}
The concept of nonorthogonality of quantum states plays a relevant
role in quantum computation and communication, cloning, and
cryptography.  Nonorthogonality is strongly related to the concept of
distinguishability, and many measures have been introduced to compare
quantum states \cite{distmeas1} and quantum processes
\cite{distmeas2}. Since the seminal work of Helstrom \cite{hel} on
quantum hypothesis testing, the problem of discriminating
nonorthogonal quantum states has received a lot of attention
\cite{rev12}. Not very much work, however, has been devoted to the
problem of discriminating general quantum operations, a part from the
case of unitary transformations \cite{CPR}. Quantum operations
describe any physically allowed transformation of quantum states,
including unitary evolutions of closed systems and non unitary
transformations of open quantum systems, such as systems interacting
with a reservoir, or subjected to noise or measurements of any
kind. The problem of discriminating quantum operations might be of
great interest in quantum error correction \cite{ec}, since knowing
which error model is the proper one influences the choice of the
coding strategy as well as the error estimation employed. Clearly,
when a repeated use of the quantum operation is allowed, a full
tomography can identify it. On the other hand, the minimum-error
discrimination approach can be useful when a restricted number of uses
of the quantum operation is considered, as in quantum hypothesis
testing \cite{hel}.

\par In this paper we consider and solve the problem of discriminating
with minimum error probability two given Pauli channels. Pauli
channels represent the most general unital channels for qubits
(e. g. bit flip, phase flip, depolarizing channels). Differently from
the case of unitary transformations \cite{CPR}, we show that
entanglement with an ancillary system at the input of the channel can
strictly improve the discrimination. We prove that an arbitrary
maximally entangled state is always an optimal input for the
discrimination, and this holds true also for generalized Pauli
channels in higher dimensional Hilbert spaces.  However, the use of
entanglement is not always needed to achieve optimality. In fact, we
compare the strategies where either entangled or unentangled states
are used at the input of the Pauli channels, and provide a necessary
and sufficient condition in terms of the structure of the channels for
which the ultimate minimum error probability can be achieved without
the need of entanglement with an ancillary system.  When such a
condition is satisfied, the optimal input state is shown to be simply
an eigenstate of one of the Pauli matrices.

\par The paper is organized as follows. In Sec. II we briefly review
the results for minimum error discrimination of two quantum states,
and formulate the problem of discrimination of two quantum operations.
In Sec. III we consider the problem for generalized Pauli channels in
the scenario where entanglement with an ancillary system is allowed at
the input of the channels.  We prove that in any dimension an
arbitrary maximally entangled state is always an optimal input for the
discrimination, and the corresponding optimal measurement is a
degenerate Bell measurement.  In Sec. IV we find the optimal strategy
for minimum error discrimination of two Pauli channels without
entanglement assistance. Finally, in Sec. V we compare the two
strategies and draw the conclusions.

\section{Discriminating quantum operations}

In the problem of discrimination two quantum states $\rho _1$ and
$\rho _2$, given with a priori probability $p_1$ and $p_2=1-p_1$,
respectively, one has to look for the two-values POVM $\{P_i \geq 0
\,, \, i=1,2\}$ with $P_1+ P_2=I $ that minimizes the error
probability
\begin{eqnarray}
p_E(P_1,P_2)
=p_1 \hbox{Tr}[\rho _1 P_2] + p_2 \hbox{Tr}[\rho_ 2 P_1]\;. 
\end{eqnarray}
We can rewrite
\begin{eqnarray}
p_E (P_1,P_2)
&=& p_1 - \hbox{Tr}[(p_1 \rho _1 -p_2 \rho_2 ) P_1]
\nonumber \\&= & 
p_2 + \hbox{Tr}[(p_1 \rho _1 -p_2 \rho_2 ) P_2]
\nonumber \\&= & 
\frac 12 \left \{
1- \hbox{Tr}[(p_1 \rho _1 -p_2 \rho_2 ) (P_1 - P_2)]\right \}
\;, 
\end{eqnarray}
where the third line can be obtained by summing and dividing the two
lines above. The minimal error probability 
\begin{eqnarray}
p_E \equiv \min_{P_1, P_2} \ p_E (P_1,P_2)
\;
\end{eqnarray}
can then be achieved by taking the orthogonal POVM $\{P_1, P_2\}$ made
by the projectors on the support of the positive and negative part of
the Hermitian operator $p_1 \rho_ 1 -p_2 \rho _2$, respectively, and
hence one has \cite{hel,locc}
\begin{eqnarray}
p_E= \frac 12 \left (1 -\Vert p_1 \rho_ 1 -p _2 \rho _2 \Vert _1
\right )\;,\label{pest}
\end{eqnarray}
where $\Vert A\Vert _1 $ denotes the trace norm of $A$. 
Equivalent expressions for the trace norm are the following \cite{bhatia} 
\begin{eqnarray}
\Vert A \Vert _1= \hbox{Tr}\sqrt{A^\dag A}=\max
  _{U}|\hbox{Tr}[UA]|= \sum _n s_n (A)\;, 
\end{eqnarray}
where the maximum is taken over all unitary operators, and $\{s_n
(A)\}$ denote the singular values of $A$. In the case of
Eq. (\ref{pest}), since the operator inside the norm is Hermitian,
the singular values just corresponds to the absolute value of the
eigenvalues.  

\par The problem of optimally discriminating two quantum operations
${\cal E}_1$ and ${\cal E}_2$ can be reformulated into the problem of
finding in the input Hilbert space $\cal H$ the state $\rho $ such
that the error probability in the discrimination of the output states
${\cal E}_1 (\rho )$ and ${\cal E}_2(\rho )$ is minimal.  We are
interested in the possibility of exploiting entanglement with an
ancillary system in order to increase the distinguishability of the
output states. In this case the output states to be discriminated will
be of the form $({\cal E}_1\otimes {\cal I}_{\cal K} ) \xi $ and
$({\cal E}_2\otimes {\cal I}_{\cal K}) \xi $, where the input $\xi $
is generally a bipartite state of ${\cal H}\otimes {\cal K}$, and the
quantum operations act just on the first party whereas the identity
map ${\cal I}={\cal I}_{\cal K}$ acts on the second.

\par In the following we will denote with $p'_E$ the minimal error
probability when a strategy without ancilla is adopted, and one has 
\begin{eqnarray}
p'_E 
=\frac 12 \left (1- \max _{\rho \in {\cal H}}\Vert p_1 {\cal E}_1
(\rho )- p_2{\cal E}_2(\rho )\Vert  _1\right )
\;.\label{peno}
\end{eqnarray}
On the other hand, by allowing the use an ancillary system, we have 
\begin{eqnarray}
p_E =\frac 12 \left (1- \max _{\xi \in {\cal H}\otimes {\cal K}}
\Vert p_1 ({\cal E}_1 \otimes {\cal I})
\xi - p_2 ({\cal E}_2\otimes {\cal I})\xi \Vert  _1\right )
\;.\label{pesi}
\end{eqnarray}
The maximum of the trace norm in Eq. (\ref{pesi}) is equivalent to the
norm of complete boundedness \cite{paulsen} of the map $p_1 {\cal
E}_1-p_2 {\cal E}_2$, and in fact for finite-dimensional Hilbert space
one can just consider $\hbox{dim}({\cal K})=\hbox{dim}({\cal H})$
\cite{paulsen,diam}. Moreover, from the linearity of quantum
operations and the convexity of the trace norm \cite{bhatia}, it
follows that in both Eqs. (\ref{peno}) and (\ref{pesi}) the maximum is
achieved by pure states.

\par Of course, $p_E \leq p_E '$. In the case of discrimination between
two unitary transformations \cite{CPR}, one has $p_E = p_E '$, namely
there is no need of entanglement with an ancillary system to achieve
the ultimate minimum error probability.

\section{Entanglement-assisted discrimination of generalized 
  Pauli channels} 
When the quantum operations can be realized from the
same set of orthogonal unitaries as random unitary transformations
\cite{orth}, namely
\begin{eqnarray}
{\cal E}_i(\rho )= \sum _n q_n ^{(i)} U_n \rho U^\dag _n\;, \qquad
\sum _n q_n^{(i)}=1 \;,\label{ds}
\end{eqnarray}
with $\hbox{Tr}[U^\dag _m U_n]=d \delta _{nm}$, and the use of an
ancillary system is allowed, one can evaluated the minimum error
probability as follows.  Let us define $r_n =p_1 q_n^{(1)}-p_2
q_n^{(2)}$. One has
\begin{eqnarray}
&&\max _ {\dket{\psi } \in {\cal H}\otimes {\cal K}}
\left \Vert p_1 ({\cal E}_1 \otimes {\cal
    I})|\psi \rangle \langle \psi |- 
p_2 ({\cal E}_2\otimes {\cal I}) |\psi \rangle \langle \psi |
\right \Vert  _1
\nonumber \\ & & =
\max _ {\dket{\psi } \in {\cal H}\otimes {\cal K}}
\left \Vert \sum _n r_n (U_n \otimes I)
  \dket{\psi }\dbra {\psi } (U^\dag _n \otimes I) \right \Vert _1
  \nonumber \\
& & \leq  \sum_n |r_n| \,
  \max _{\dket{\psi } \in {\cal H}\otimes {\cal K}}
\left \Vert (U_n \otimes I)
  \dket{\psi }\dbra {\psi } (U^\dag _n \otimes I) \right \Vert _1
=\sum_n |r_n| \;.\label{20}
\end{eqnarray}
The bound is Eq. (\ref{20}) can be saturated by the maximally
entangled state
\begin{eqnarray}
|\Psi \rangle =\frac{1}{\sqrt d}\sum _{n=0}^{d-1}
{|n \rangle |n \rangle }\;
\end{eqnarray}
as we show in the following. Let us define 
\begin{eqnarray}
A\equiv \sum _n r_n (U_n \otimes I)\dket{\Psi }\dbra{\Psi }(U_n ^\dag
\otimes I)
\;\label{aaa}
\end{eqnarray}
Notice that 
\begin{eqnarray}
\dbra{\Psi } (U^\dag _n \otimes I) (U_m \otimes I) \dket{\Psi}=\frac
1d \Tr [U^\dag _m U_n]=\delta _{nm} \;,\label{}
\end{eqnarray}
namely $A$ is diagonal on maximally entangled states. Then one has 
\begin{eqnarray}
\Vert A \Vert _1 = \Tr \sqrt{A^\dag A}= \sum _n |r_n|\;.
\end{eqnarray}
It follows that 
\begin{eqnarray}
p_E=\frac 12 \left(1- \sum_n |r_n|\right )\;.\label{pepc} 
\end{eqnarray}
The corresponding measurement to be performed at the output of the
channel is given by the projectors on support of the positive and
negative part of the operator $A$ in Eq. (\ref{aaa}), namely  
\begin{eqnarray}
&&P_1 = 
\sum _ {n_+}  (U_{n_+} \otimes I )
|\Psi \rangle \langle \Psi | (U^\dag _{n_+} \otimes I )\;,
\nonumber \\& & 
P_2 = \sum _ {n_-}  (U_{n_-} \otimes I )
|\Psi \rangle \langle \Psi | (U^\dag _{n_-} \otimes I )\;,
\label{pii}
\end{eqnarray}
where the index $n_+$ ($n_-$) are in correspondence with the positive
(negative) elements of $\{r_n \}$. Notice that the set of projectors
$\{(U_{n} \otimes I ) |\Psi \rangle \langle \Psi | (U^\dag _{n}
\otimes I )\}$ are orthogonal maximally entangled states, and hence
the measurement is a degenerate Bell measurement \cite{pla}.  For the
unitarily invariance property of the trace norm \cite{bhatia}, the
minimal error probability can always be achieved by using any {\em
arbitrary} maximally entangled state at the input, namely $(I\otimes
V) |\Psi \rangle $ with $V$ unitary.  The corresponding optimal
measurement will be $\{(I\otimes V) P_i (I \otimes V^\dag )\}$, with
$\{P_i \}$ as in Eq.  (\ref{pii}).  \par By dropping the condition of
orthogonality of the $\{U_n \}$, one just obtains the bounds
\begin{eqnarray}
\frac 12 \left(1- \sum_n |r_n| 
\right )
\leq p_E \leq 
\frac 12 \left(1- \Vert A \Vert _1 \right ) \;,  
\end{eqnarray}
since Eq. (\ref{20}) gives the lower bound, whereas the upper bound is
simply obtained by taking as input the maximally entangled state.  
\section{Discrimination of Pauli channels with  no  entanglement 
assistance} \label{sec4} 
In this section we consider the case of
discrimination of two Pauli channels for qubits, namely
\begin{eqnarray}
{\cal E}^{(1)}(\rho )= \sum_{n  =0}^3 q_n  ^{(1) }\sigma
  _n  \rho \sigma _n  \;, \qquad 
{\cal E}^{(2)}(\rho )= \sum_{n  =0}^3 q_n  ^{(2) }\sigma
  _n  \rho \sigma _n  \;,\label{chan2}
\end{eqnarray}
where $\{\sigma _0\,, \sigma _1\,,\sigma _2\,,\sigma _3 \}= \{I\,,
\sigma _x\,,\sigma _y\,,\sigma _z\}$ and $\sum _{n =0}^3 q_n ^{(1)} =
\sum _{n =0}^3 q_n ^{(2)} = 1$. As shown in the previous section, the
minimal error probability when entanglement with an ancillary system
is used at the input is given by Eq. (\ref{pepc}), where 
\begin{eqnarray}
r_n = p_1
q_n ^{(1)}-p_2 q_n ^{(2)} 
\;,\label{renne}
\end{eqnarray}
and $p_1$ and $p_2$ denote the a priori
probabilities.

Here, we are interested to understand when the
entangled-input strategy is really needed to achieve the optimal
discrimination, hence we derive in the following 
the optimal strategy with no ancillary system. 
According to Eq. (\ref{peno}) the minimal error probability is given
by 
\begin{eqnarray}
p'_E= \frac 12 \left (1 - \max _{|\psi \rangle \in {\cal H}} \left
\Vert \sum _{n=0}^3 
r_ n \sigma _n \dket{\psi }\dbra {\psi }\sigma _n \right
\Vert _1 \right )
\end{eqnarray}
By parameterizing the pure state of the qubit as 
\begin{eqnarray}
\dket{\psi }=\cos \frac \theta 2 |0 \rangle + e^{i \phi }\sin \frac \theta 2 |1
\rangle \;,
\end{eqnarray}
one has 
\begin{eqnarray}
&&\!\!\!\!\!\!\!\!\!\!\!\!\!\!\!\!\!\!\!\!\!\!\!\!\!\!\!\!\!\!\!\! 
\xi \equiv
\sum
  _{n=0}^3  
r_ n   \sigma _n   \dket{\psi }\dbra {\psi }\sigma
  _n  =  \\& & 
\!\!\!\!\!\!\!\!\!\!\!\!\!\!\!\!\!\!\!\!\!\!\!\!\!\!\!\!\!\!\!\! 
\left(
\begin{array}{cc}
(r_0+r_3) \cos ^2 \frac \theta 2 + (r_1 +r_2) \sin ^2 \frac \theta 2& 
\frac 1 2 \sin \theta   
[(r_0 -r_3)e^{i\phi }+(r_1 -r_2)e^{-i\phi }]
\\
\frac 1 2 \sin \theta  
[(r_0 -r_3)e^{-i\phi }+(r_1
  -r_2)e^{i\phi }]
& (r_0+r_3) \sin ^2 \frac \theta 2 + (r_1 +r_2) \cos ^2 \frac \theta 2
\end{array}
\right)\;.
\nonumber 
\end{eqnarray}
The eigenvalues of $\xi $ are given by 
\begin{eqnarray}
&&\!\!\!\!\!\!\!\!\!\!\!\!\!\!\!\!\!\!\!\!\!\!\!\!
\lambda  (\theta ,\phi )_{1,2}= \frac 12 \left\{ r_0+r_1+r_2+r_3 \pm  
\left [ \cos ^2 \theta [r_0+r_3 - r_1 -r_2]^2  \right. \right.
\nonumber \\& & 
\!\!\!\!\!\!\!\!\!\!\!\!\!\!\!\!\!\!\!\!\!\!\!\!
\left. \left. + \sin ^2 \theta
  [(r_0- r_3)^2  +(r_1-r_2)^2
+2 \cos (2\phi )
(r_0-r_3)(r_1-r_2)] \right ] ^{1/2} \right \}\;.
\end{eqnarray}
We then have 
\begin{eqnarray}
p'_E= \frac 12 \left (1 - \max _{\theta, \phi  } [|\lambda _1 (\theta
  ,\phi )|+ |\lambda _2 (\theta ,\phi )|] \right  )\;.\label{21}
\end{eqnarray}
Notice that the function $f(\theta ,\phi)\equiv (|\lambda _1 (\theta
  ,\phi )|+ |\lambda _2 (\theta ,\phi )|)  $ can be rewritten as  
\begin{eqnarray}
\!\!\!\!\!\!\!\!\!\!\!\!\!\!\!\!\!\!\!\!
f(\theta ,\phi)= &&\max \{ |r_0+r_1 + r_2 +r_3|\,, 
\{\cos ^2 \theta [r_0+r_3 - r_1 -r_2]^2 
\nonumber \\& & 
+\sin^2 \theta
  [(r_1 -r_2)^2+(r_0- r_3)^2  +2 \cos (2\phi)
(r_0-r_3)(r_1-r_2)]\}^{1/2}\} 
\;.
\end{eqnarray}
The maximum over $\theta ,\phi $ in Eq. (\ref{21}) 
can be found just by comparing the
values of $f(\theta ,\phi )$ at the stationary points, namely $\theta
= k \pi $, and $\theta = \frac {(2k+1)\pi}{2} 
\,, \phi = l\frac \pi 2$,
with $k,l$ integer. 
Since one has 
\begin{eqnarray}
&&\!\!\!\!\!\!\!\!\!\!\!\!\!\!\!\!\!\!\!\!\!\!\!\!\!\!\!\!\!\!\!\! 
\ \ \ \ 
2\left [ 
|\lambda _1 (k \pi ,
  ,\phi )|+ |\lambda _2 (k\pi ,\phi )| \right]\nonumber \\ &&
\!\!\!\!\!\!\!\!\!\!\!\!\!\!\!\!\!\!\!\!\!\!\!\!\!\!\!\!\!\!\!\! 
=|r_0+r_1 + r_2 +r_3 +
|r_0 +r_3 - r_1 -r_2 ||+
|r_0+r_1 + r_2 +r_3 -
|r_0 +r_3 - r_1 -r_2
||
\nonumber \\ &&
\!\!\!\!\!\!\!\!\!\!\!\!\!\!\!\!\!\!\!\!\!\!\!\!\!\!\!\!\!\!\!\! 
=2(|r_0+r_3| + |r_1 +r_2|)\;;
\end{eqnarray}
\begin{eqnarray}
& & \!\!\!\!\!\!\!\!\!\!\!\!\!\!\!\!\!\!\!\!\!\!\!\!\!\!\!\!\!\!\!\! 
\ \ \ \ 
2\left [ \left |\lambda _1 \left (\frac {(2k+1)\pi}{2} ,
  l \pi  \right )\right |+ \left |\lambda _2 \left (
\frac{(2k+1)\pi}{2} , 
l\pi  \right )\right | \right ]
\nonumber \\&&
\!\!\!\!\!\!\!\!\!\!\!\!\!\!\!\!\!\!\!\!\!\!\!\!\!\!\!\!\!\!\!\! 
= 
|r_0+r_1 + r_2 +r_3 +
|r_0 -r_3 +r_1 -r_2 ||+
|r_0+r_1 + r_2 +r_3 -
|r_0 -r_3 +r_1 -r_2 ||
\nonumber \\&&
\!\!\!\!\!\!\!\!\!\!\!\!\!\!\!\!\!\!\!\!\!\!\!\!\!\!\!\!\!\!\!\! 
=2(|r_0+r_1| + |r_2 +r_3|)\;;
\end{eqnarray}
\begin{eqnarray}
& & 
\!\!\!\!\!\!\!\!\!\!\!\!\!\!\!\!\!\!\!\!\!\!\!\!\!\!\!\!\!\!\!\! 
\ \ \ \ 
2\left [\left |\lambda _1 \left (\frac {(2k+1)\pi}{2} ,
  \frac {l \pi}{2}  \right )\right |+ \left |\lambda _2 \left 
(\frac {(2k+1)\pi}{2} ,
  \frac{l\pi}{2}  \right )
\right | \right]
\nonumber \\&&
\!\!\!\!\!\!\!\!\!\!\!\!\!\!\!\!\!\!\!\!\!\!\!\!\!\!\!\!\!\!\!\! 
=
|r_0+r_1 + r_2 +r_3 +
|r_0 -r_3 - r_1  +r_2 ||+
|r_0+r_1 + r_2 +r_3 -
|r_0 -r_3 -r_1 + r_2 ||
\nonumber \\&&
\!\!\!\!\!\!\!\!\!\!\!\!\!\!\!\!\!\!\!\!\!\!\!\!\!\!\!\!\!\!\!\! 
=2(|r_0+r_2| + |r_1 +r_3|)
\;;
\end{eqnarray}
one finally obtains 
\begin{eqnarray}
\!\!\!\!\!\!\!\!\!\!\!\!\!\!\!\!\!\!\!\!p_E '=\frac 12 \left (1 - M \right )
\;,\label{pem} 
\end{eqnarray}
where 
\begin{eqnarray}
\!\!\!\!\!\!\!\!\!\!\!\!\!\!\!\!\!\!\!\!
M= \max \left\{ |r_0+ r_3|+|r_1+r_2|\,, |r_0+ r_1|+|r_2+r_3|\,, |r_0+
r_2|+|r_1+r_3| \right \} \;.\label{emme}
\end{eqnarray}
The three cases inside the brackets corresponds to using an
eigenstate of $\sigma _z$, $\sigma _x$, and $\sigma _y$, respectively,
as input state of the unknown channel.  The corresponding measurements
to be performed at the output are the three Pauli matrices
themselves. 
\section{Conclusion}
From comparing Eqs. (\ref{pem}) and (\ref{emme}) with
Eq. (\ref{pepc}), one can see that entanglement with an ancillary
system is not needed to achieve the ultimate minimal error probability
in the discrimination of the two Pauli channels of Eq. (\ref{chan2})
as long as $M=\sum _{n=0}^3 |r_n|$, with $r_n$ given in
Eq. (\ref{renne}).  This happens when $\Pi _{n=0}^3 r_n \geq 0$.
Hence, entanglement assistance is necessary if and only if all
$\{r_n\}$ are different from zero, with three of them with the same
sign, and the remaining one with the opposite sign.  Among these
cases, there are striking examples where the channels can be perfectly
discriminated only by means of entanglement. This is the case of two
channels of the form
\begin{eqnarray}
{\cal E}_1(\rho )=\sum _{n  \neq m }q_n   \sigma _n 
  \rho \sigma _n  \;,\qquad 
{\cal E}_2(\rho )=\sigma _m  \rho \sigma _m
\;, 
\end{eqnarray}
with $q_n \neq 0$, and arbitrary a priori probability. This example
can be simply understood, since the entanglement-assisted strategy
increases the dimension of the Hilbert space such that the two
possible output states will have orthogonal support.

In conclusion, we considered the problem of discriminating two Pauli
channels with minimal error probability. We showed that using
maximally entangled states with an ancillary system at the input of
the channel allows to achieve the optimal discrimination, and this
holds true also for generalized Pauli channels in higher dimensional
Hilbert spaces.  In the case of qubits, we also found the minimal
error probability for the discrimination strategy with no entanglement
assistance, and showed that the optimal input states are the
eigenstates of one of the Pauli matrices. By comparison, we then
characterized in a simple way the instances where the optimal
discrimination can be achieved without the need of entanglement.  

\par It could be interesting to look for similar conditions in the
case of generalized Pauli channels in higher dimension. For this
problem, one should translate the algebraic derivation in
Sec. \ref{sec4} into a more geometrical picture. This could result in
better physical insights into why in some cases entanglement is not
required to achieve minimum error discrimination. As in the field of
state discrimination, one could also study the problem of optimal
unambiguous discrimination \cite{unam} of channels, where the
unambiguity is paid by the possibility of getting inconclusive results
from the measurement.  Finally, an alternative approach is to consider
the problem in the frequentistic scenario, instead of the
Bayesian one, as it has been recently studied for state
discrimination \cite{kahn}. In this case, one does not have a priori
probabilities and has to maximize the worst probability of correct
detection.

\ack This work has been sponsored by INFM through the
project PRA-2002-CLON, and by EC and MIUR through the cosponsored
ATESIT project IST-2000-29681 and Cofinanziamento 2003.


\begin{thebibliography}{99}
\bibitem{orth}This generalizes Pauli channels in arbitrary
dimension. An example of such a set is operators of the form $X^kZ^l$
where $X$ and $Z$ act on the basis states $|0 \rangle , ...,|d-1
\rangle $ as $X| j \rangle =|j \oplus 1 \rangle $ and $Z |j \rangle
=e^{2\pi i j/d}|j \rangle $, with $\oplus $ denoting addition modulo 
$d$.
\end{thebibliography}
\end{document}